\documentclass[prd,preprint,tightenlines,floatfix,noshowpacs,preprintnumbers,nofootinbib,eqsecnum,superscriptaddress]{revtex4-1}

\usepackage[dvips,final]{graphicx}
\usepackage{amssymb}
\usepackage{amsmath}
\usepackage{amsfonts}
\usepackage{epsfig}
\usepackage{bm}
\usepackage[section]{placeins}
\usepackage{multirow}
\usepackage{ctable}
\usepackage{booktabs}
\usepackage{array}
\usepackage{tabularx}
\usepackage{xcolor}
\usepackage{pstricks}
\usepackage{xspace}

\newcommand{\bp}{\mbox{\boldmath $p$}}

\newcommand{\bk}{\mbox{\boldmath $k$}}

\newcommand{\kt}{\ensuremath{k_{\mathrm{T}}}\xspace}
\newcommand{\der}{\ensuremath{\mathrm{d}}\xspace}

\begin{document}

\vfill
\title{Production of $t \bar t$ pairs
via $\gamma \gamma$ fusion \\ with photon transverse momenta
and proton dissociation}

\author{Marta {\L}uszczak}
\email{luszczak@ur.edu.pl}
\affiliation{
Faculty of Mathematics and Natural Sciences,
University of Rzesz\'ow, ul. Pigonia 1, PL-35-310 Rzesz\'ow, Poland}

\author{Laurent Forthomme}
\email{laurent.forthomme@cern.ch}
\affiliation{Helsinki Institute of Physics,
P.O.Box 64 (Gustaf H{\"{a}}llstr{\"{o}}min katu 2),
FI-00014 University of Helsinki, Helsinki, Finland}

\author{Wolfgang Sch\"afer}
\email{wolfgang.schafer@ifj.edu.pl}
\affiliation{Institute of Nuclear Physics Polish Academy of Sciences,
ul. Radzikowskiego 152, PL-31-342 Krak\'{o}w, Poland}

\author{Antoni Szczurek}
\email{antoni.szczurek@ifj.edu.pl}
\affiliation{Institute of Nuclear Physics Polish Academy of Sciences,
ul. Radzikowskiego 152, PL-31-342 Krak\'{o}w, Poland}

\date{\today}

\begin{abstract}
We discuss the production of $t \bar t$ quark-antiquark pairs in
proton-proton collisions via the $\gamma \gamma$ fusion mechanism.
We include topologies in which both protons stay intact or one or even both
of them undergo dissociation. The calculations are
performed within the $k_{\rm T}$-factorisation approach, including transverse
momenta of intermediate photons. Photon fluxes associated to
inelastic (dissociative) processes are calculated based on modern
parameterisations of proton structure functions.
We find an integrated cross section of about 2.36 fb at $\sqrt{s}$ = 13 TeV for all contributions.
The cross section for the fully elastic process is the smallest.
Inelastic contributions are significantly
reduced when a veto on outgoing jets is imposed.
We present several differential distributions in rapidity and transverse
momenta of single $t$ or $\bar t$ quarks/antiquarks as well as distributions in
invariant mass of both the $t \bar t$ and masses of dissociated
systems. A few two-dimensional distributions are presented in addition.
\end{abstract}

\pacs{}


\maketitle

\section{Introduction}

Photon-induced processes in proton-proton or nucleus-nucleus
interactions have become very topical recently. The large energy at the LHC,
when combined with relatively large luminosity at run II, allows
to start the exploration of such processes. Although in the case
of proton-proton collisions the relevant cross sections
are rather small, some of their salient features may
allow their measurement.
For instance, photon-fusion events will lead to a
rapidity gap observable experimentally between the electromagnetic/electroweak
vertex and forward scattered systems.

The production of electron or muon pairs are flag examples.
Recently both CMS \cite{Chatrchyan:2011ci,Chatrchyan:2012tv,Cms:2018het}
and ATLAS \cite{Aad:2015bwa} studied
such processes. Another example is the production of $W^+ W^-$ pairs
also studied by both Collaborations
\cite{Chatrchyan:2013akv,Khachatryan:2016mud,Aaboud:2016dkv}.
These results allow to obtain upper limits on the deviations from
Standard Model couplings.
On the theoretical side such processes may be calculated using the
equivalent photon approximation for purely elastic and partially or
fully inelastic processes \cite{Budnev:1974de}. The photon flux corresponding to
the elastic part is then expressed in terms of electromagnetic
form factors of proton (electric and magnetic, or equivalently Dirac and Pauli).
Proton dissociative processes need as an input the structure functions $F_2$ and $F_L$
of a proton, here especially $F_2$ is well known in a broad kinematic range
from a large body of deep-inelastic electron-proton scattering data.

A similar method was used in the \kt-factorisation approach for dilepton
production \cite{daSilveira:2014jla,Luszczak:2015aoa} and recently
for $W^+ W^-$ production \cite{Luszczak:2018ntp}.
Actually a similar approach was suggested for lepton pairs
long time ago in \cite{Vermaseren:1982cz} and realised in
the LPAIR code \cite{Baranov:1991yq}.

Different parameterisations of the proton structure functions were
used in the literature. The overall errors/uncertainties
are therefore associated with insufficient knowledge of structure functions
and/or poor functional form of parameterising the data.
In \cite{Manohar:2016nzj,Manohar:2017eqh} it was argued that
parameterisations based on proton structure functions
have much smaller uncertainties, and
generally lead to much smaller
cross sections
than in
the standard DGLAP approach.


In the present approach we study a new final state, namely $t$ and $\bar t$. Being the heaviest of fundamental
Standard Model particles,
the top quark is of special interest. The dominant production
mechanisms investigated until recently in great detail involve the
strong interactions.
While the precision studies of electroweak production
mechanisms are clearly the task for an $e^+ e^-$ collider
\cite{Abramowicz:2018rjq}, here we wish to investigate the
$\gamma \gamma$ fusion contribution in $pp$ collisions at the LHC.

There has not been much discussion of this final state in
the literature in the context of the $\gamma \gamma$ fusion.
In Ref.\cite{Fayazbakhsh:2015xba}, the fully exclusive
$pp \to pp t \bar t$ process was discussed at LHC energies, including
possible anomalous $\gamma t \bar t$ couplings.
A very comprehensive study \cite{Czakon:2017wor} includes electroweak
corrections to inclusive $t \bar t$ production.
Being a part of these corrections, the $\gamma \gamma$
fusion subprocess is evaluated using collinear photon
parton distributions. The $\gamma \gamma$ contribution to inclusive
$t \bar t$ production is found to be negligible, when realistic photon
distributions are used.

\begin{figure}
  \centering
  \includegraphics[width=.325\textwidth]{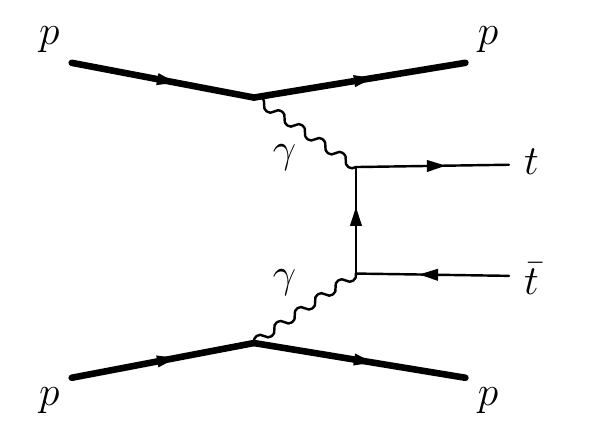}
  \includegraphics[width=.325\textwidth]{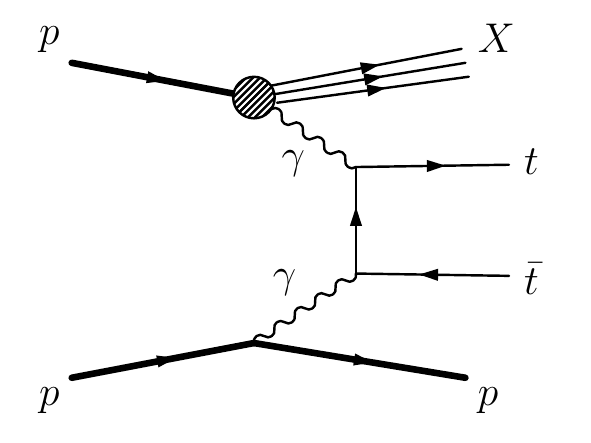}
  \includegraphics[width=.325\textwidth]{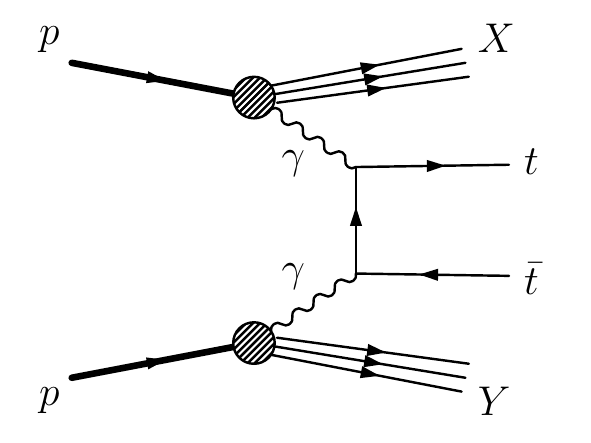}
  \caption{Classes of processes discussed in the present
paper. From left to right: elastic-elastic, inelastic-elastic (or equivalently, elastic-inelastic), and inelastic-inelastic contributions.}
\label{fig:diagrams}
\end{figure}

In Fig.\ref{fig:diagrams} we show diagrams of the four different classes
of processes included in our present analysis.
In the present paper we concentrate on general characteristics and study
of differential distribution to select a proper observable for
future experimental studies.

We wish to evaluate the cross section separately for each category presented
in the figure within the Standard Model. We also aim at calculating
several differential distributions of interest.

\section{A sketch of the formalism}
Our calculations are based on unintegrated photon fluxes, which depend on the longitudinal
momentum fraction, $x$ and the transverse momentum (in the $pp$-frame) $\bk$ of the photon,
and on the invariant mass $M_X$ of the dissociative system in the $p \to \gamma X$ vertex:
\begin{eqnarray}
\frac{d \gamma (x,\bk,M_X)}{d M_X} = \gamma_{\rm el}(x, \bk) \delta(M_X - m_p)
+  \frac{d \gamma_{\rm inel}(x,\bk,M_X)}{d M_X}
\theta\big(M_X -(m_p + m_\pi)\big),
\nonumber \\
\end{eqnarray}
Here we put in evidence the contributions of ``elastic'' processes
with an intact proton in the final state as well as the ``inelastic'' component
for hadronic final states $X \neq p$.


\begin{eqnarray}
\frac{d \sigma(pp \to X (\gamma^\ast \gamma^\ast \to t \bar t) Y)}{d y_+ d y_- d^2\bp_{\rm T1} \der^2\bp_{\rm T2} d M_X d M_Y} =
 \int d^2 \bk_1 d^2 \bk_2
\frac{x_1 d \gamma(x_1,\bk_1,M_X)}{d M_X}
\frac{x_2 d \gamma(x_2,\bk_2,M_Y)}{d M_Y} \; \times\hspace{3em}\nonumber \\
\times \; \frac{1}{16 \pi^2 (x_1 x_2 s)^2}
|M(\gamma^* \gamma^* \to t \bar t; \bk_1,\bk_2)|^2 \,
\delta^{(2)}(\bp_{\rm T1} + \bp_{\rm T2} - \bk_1 - \bk_2 ) . \nonumber \\
\end{eqnarray}
In this formalism $y_\pm$ are the rapidities and $\bp_{\rm T1,2}$ the transverse momenta of the $t$ and $\bar t$
quark respectively. The off-shell matrix element squared
for the $\gamma^* \gamma^* \to t \bar t$ process is the same
as the one for dilepton production found in \cite{daSilveira:2014jla}, up to a factor
$e_f^4 N_c$, with $e_f = +2/3$ the quark electric charge, and $N_c =3$. In results shown below we use the top quark mass $m_t = 173 \, {\rm GeV}$.

These formulas are implemented in CepGen \cite{Forthomme:2018ecc} for the
Monte-Carlo generation of unweighted events containing the $t \bar t$ pair as well as
remnants of mass $M_X, M_Y$.

Details on the relation between photon fluxes and proton structure functions
may be found in \cite{Luszczak:2018ntp}, where we also describe several parameterisations
used for $F_2, F_L$.
Here we only mention that in the region of $Q^2>9 \, {\rm GeV}^2$, which is the most
important one for the process at hand, we use a
perturbative QCD NNLO calculation of \cite{Martin:2009iq}.

\section{Results}

In Table \ref{tab:sig_tot} we show integrated cross sections for each
of the categories of $\gamma \gamma$ processes shown in Fig.\ref{fig:diagrams}.
We observe the following hierarchy as far as the integrated
cross section is considered:
\begin{equation}
\sigma_{t \bar t}^{\rm el-el} < \sigma_{t \bar t}^{\rm in-el}
= \sigma_{t \bar t}^{\rm el-in} < \sigma_{t \bar t}^{\rm in-in}.
\label{eq:sigma_hierarchy}
\end{equation}
The summed cross section at $\sqrt{s} = 13 \, \rm{TeV}$
is 2.36 fb. This is a rather small number in comparison with other inclusive production mechanisms.
A possible extraction of $\gamma \gamma$ events therefore requires, e.g.
experimental cuts on rapidity gaps.
So far we have ignored the gap survival factor due to remnant
fragmentation and/or soft processes. However such effects may reverse the order
of Eq.~\ref{eq:sigma_hierarchy}.
This behaviour was obtained previously for production of $W^+ W^-$ pairs
via $\gamma-\gamma$ fusion in \cite{Forthomme:2018sxa}.
In the right panel of Table \ref{tab:sig_tot} we show results when an extra condition on the
rapidity of the recoiling jet, -2.5 $< y_{\rm jet} <$ 2.5, is imposed. The condition on jet rapidity
gives very similar results as a condition on charged particles \cite{Forthomme:2018sxa}
when including explicitly remnant fragmentation.

\begin{figure}
  \includegraphics[width=.60\textwidth]{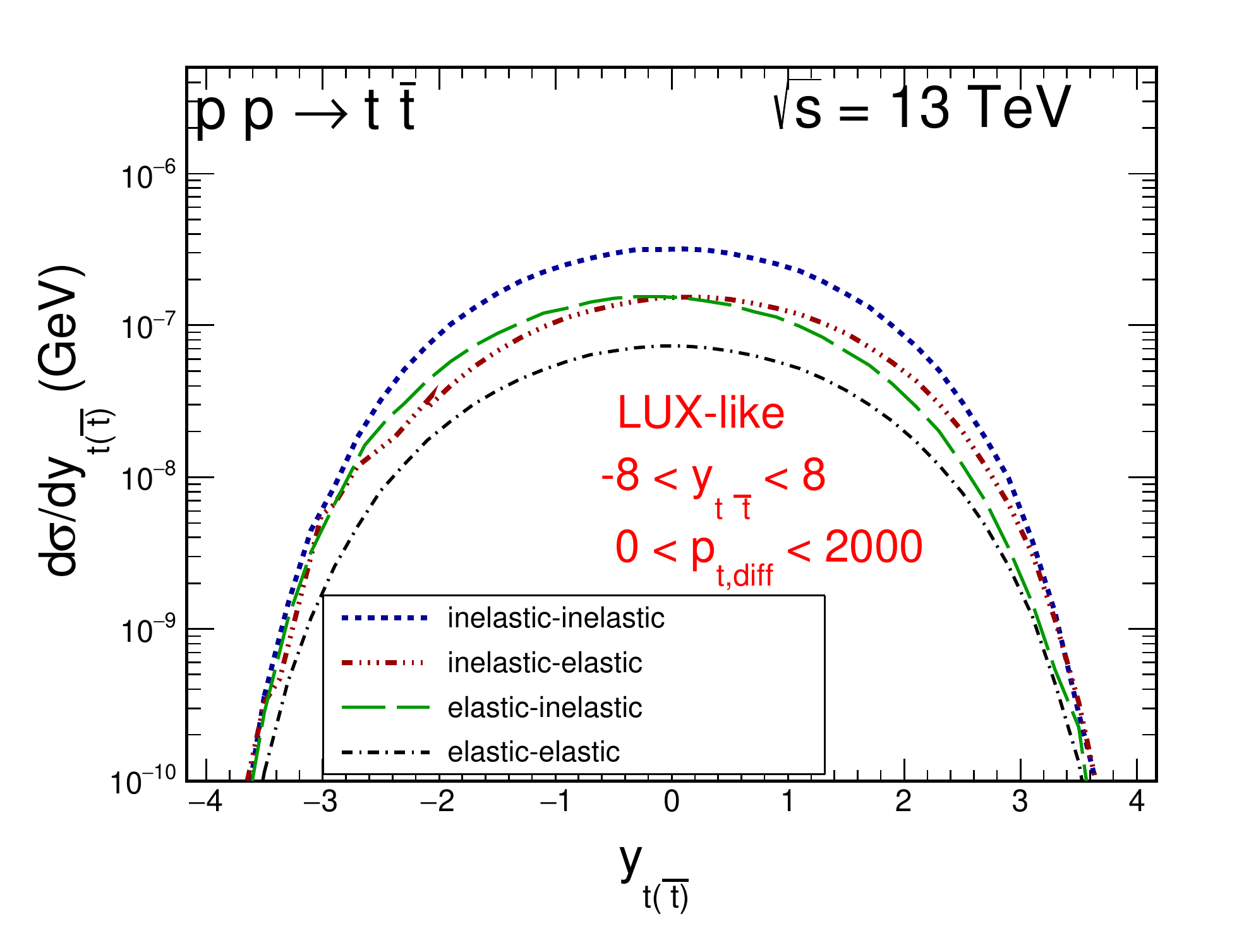}
  \caption{Rapidity distribution for different components defined
in the figure.
}
\label{fig:dsig_dy}
\end{figure}

\begin{table}[tbp]
\begin{tabular}{|c|c|c|}
\hline
Contribution          &  No cuts & $y_{\rm jet}$ cut\\
\hline
elastic-elastic       &  0.292 &  0.292 \\
elastic-inelastic     &  \multirow{2}{*}{0.544} &  \multirow{2}{*}{0.439} \\
inelastic-elastic     &  & \\
inelastic-inelastic   &  0.983 &  0.622 \\
\hline
all contributions     &  2.36  &  1.79 \\
\hline
\end{tabular}
\caption{Cross section in fb at $\sqrt{s}$ = 13 TeV for
different components (left column) and the same when the extra condition
on the outgoing jet -2.5 $< y_{\rm jet} <$ 2.5 is imposed.}
\label{tab:sig_tot}
\end{table}
In Fig.\ref{fig:dsig_dy} we show the rapidity distributions of $t$ quarks
or $\bar t$ antiquarks (these are identical) for different categories of the final
state. Quite similar distributions are obtained for the different
categories of processes. There is only a small asymmetry with respect to
$y$ = 0 for elastic-inelastic or inelastic-elastic contributions.
By construction, the sum of both the contributions is symmetric with respect to $y=0$.
The cross section is concentrated at intermediate rapidities so
in principle should be measurable by the ATLAS/CMS central detectors.
However, a precise estimation would require imposing cuts on the decay products of $t$ and $\bar t$
(for instance, into a $b$ jet and a charged lepton). This goes beyond scope
and aim of the present work.

\begin{figure}
  \includegraphics[width=.60\textwidth]{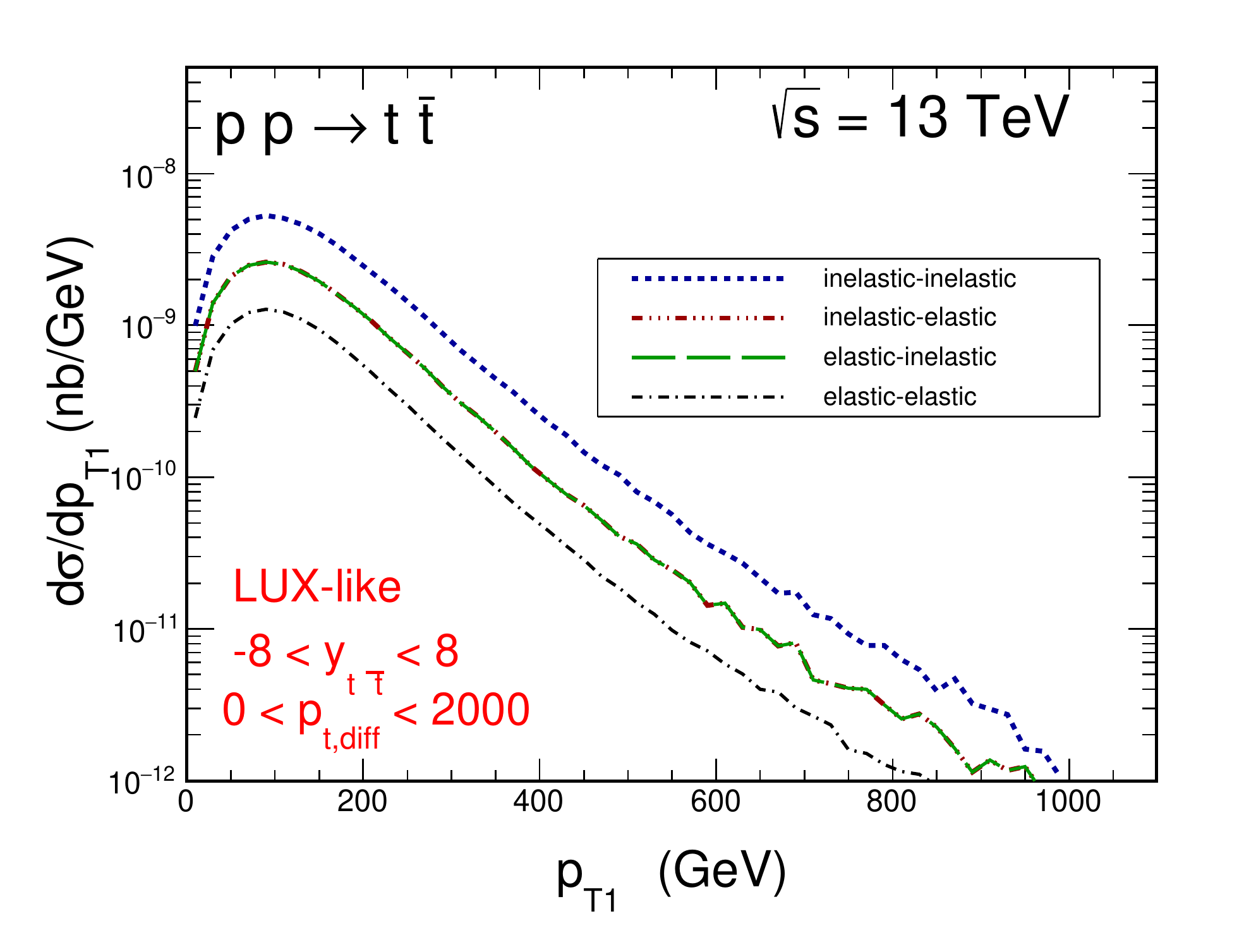}
  \caption{Transverse momentum distribution of $t$ or $\bar t$
for different components defined in the figure.
}
\label{fig:dsig_dpt}
\end{figure}

In Fig.\ref{fig:dsig_dpt} we show the distribution in transverse momentum
of $t$ or $\bar t$ (identical). Here again the different categories give
distributions of similar shape.

\begin{figure}
  \includegraphics[width=.48\textwidth]{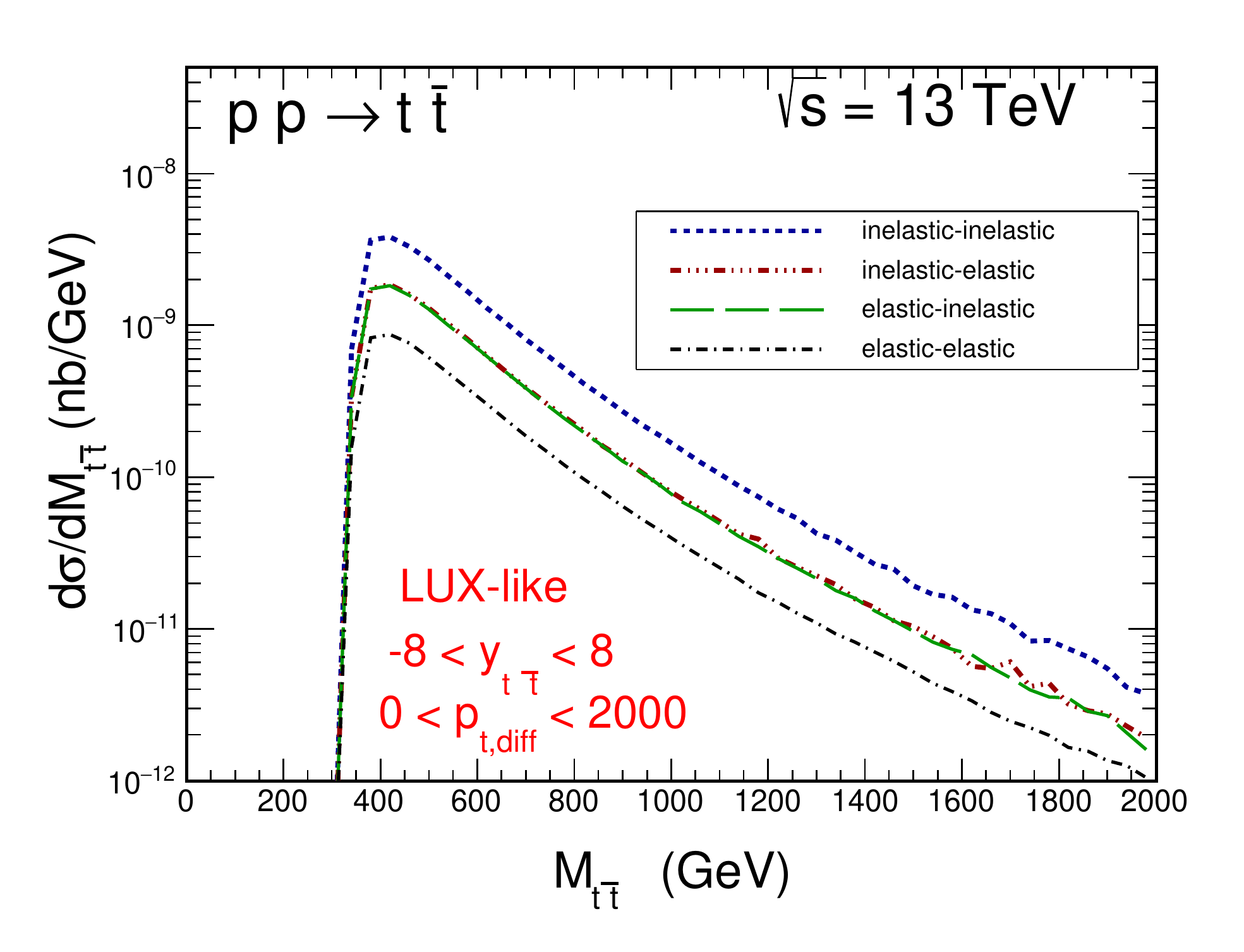}
  \includegraphics[width=.48\textwidth]{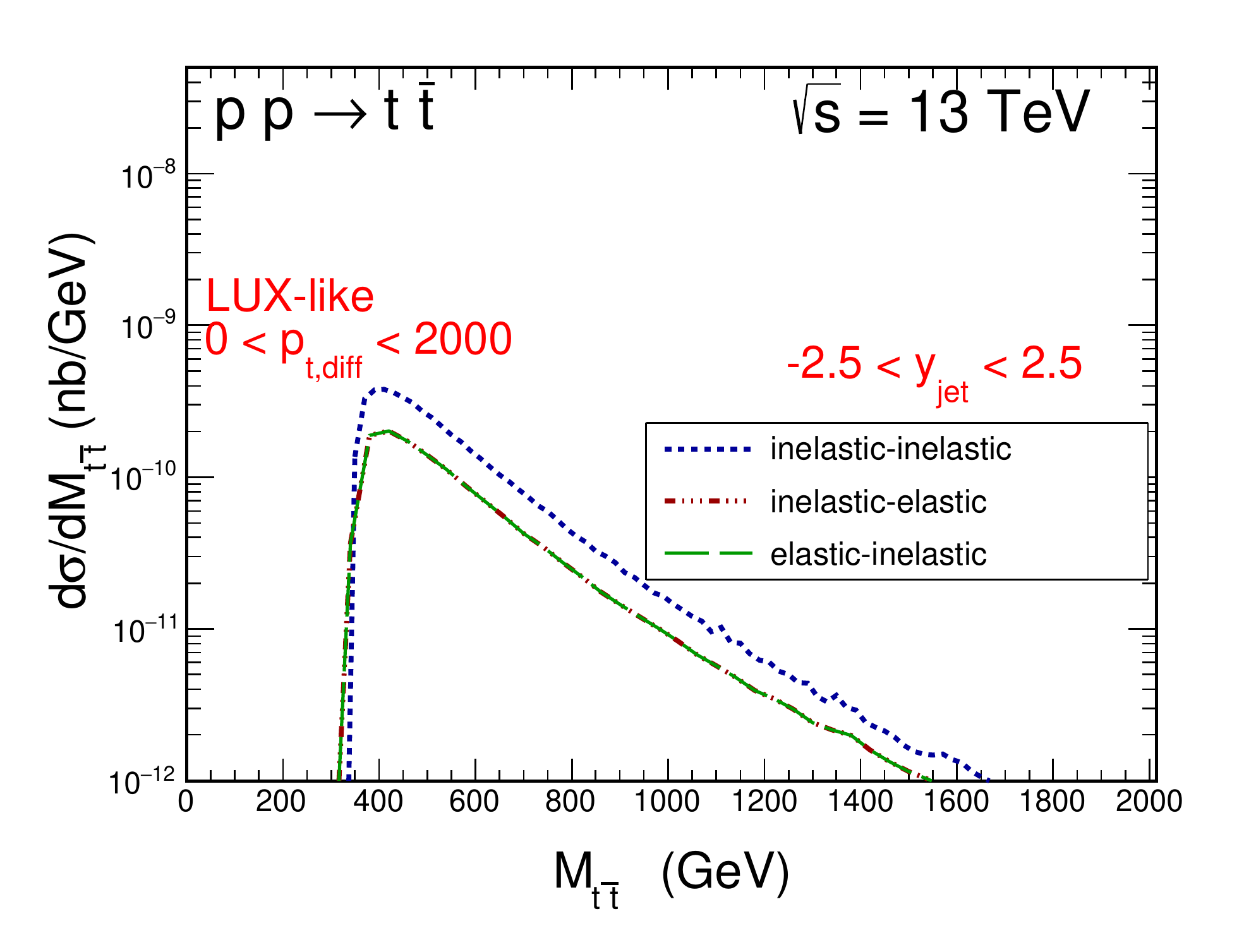}
  \caption{$t \bar t$ invariant mass distribution for different
components defined in the figure. The left panel is without imposing
the condition on the struck quark/antiquark and the right panel includes
the condition.
}
\label{fig:dsig_dMttbar}
\end{figure}

The same is true for the distribution in $t \bar t$ invariant mass
(see the left panel of Fig.\ref{fig:dsig_dMttbar}). The distributions
are almost identical and differ only by normalisation.
For completeness in the right panel of Fig.\ref{fig:dsig_dMttbar} we show
similar results when conditions on outgoing light
quark/antiquark jets are imposed.
The extra condition leads to a
lowering of the cross section with only very small modification of
the shape in $M_{t \bar t}$.

\begin{figure}
  \includegraphics[width=.48\textwidth]{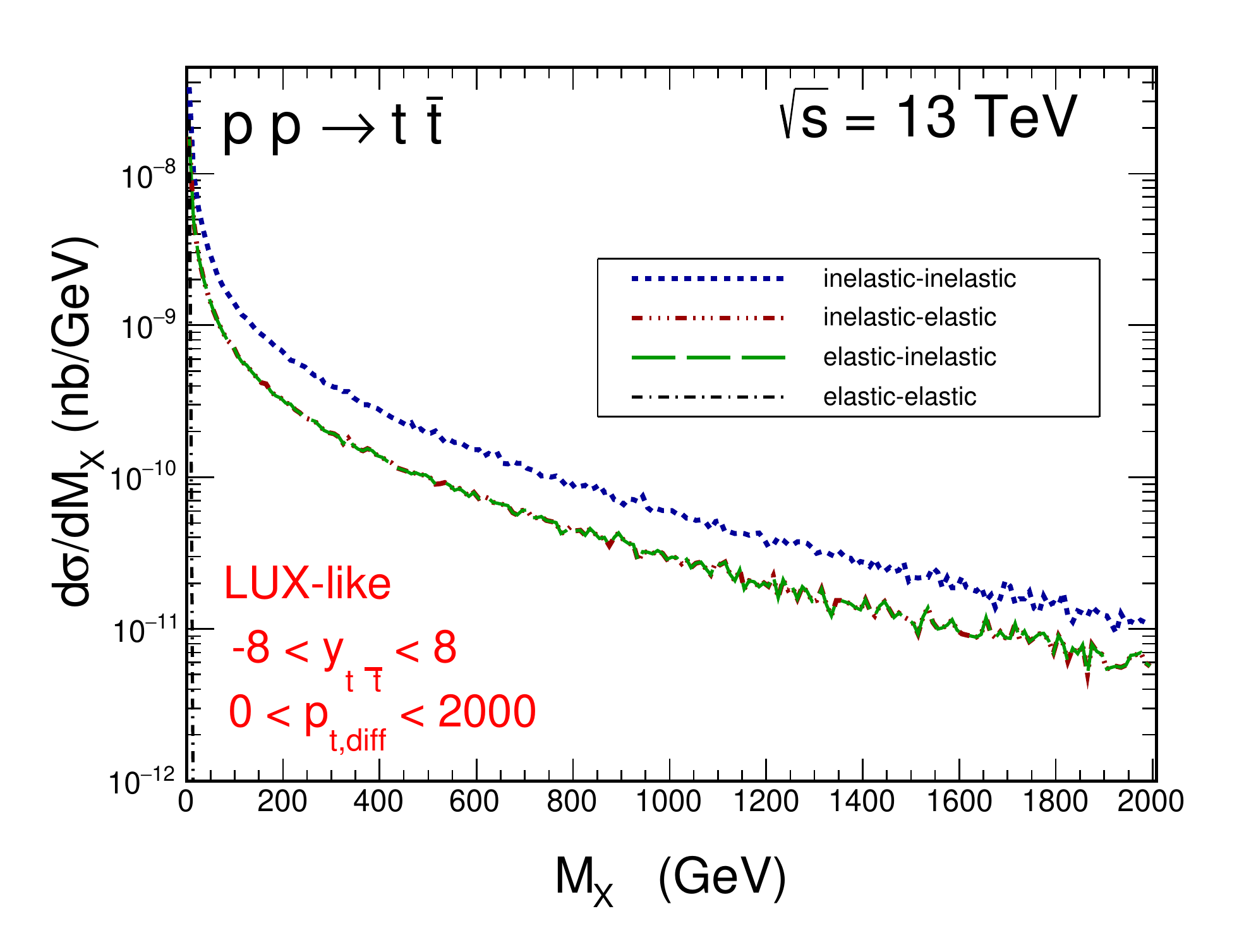}
  \includegraphics[width=.48\textwidth]{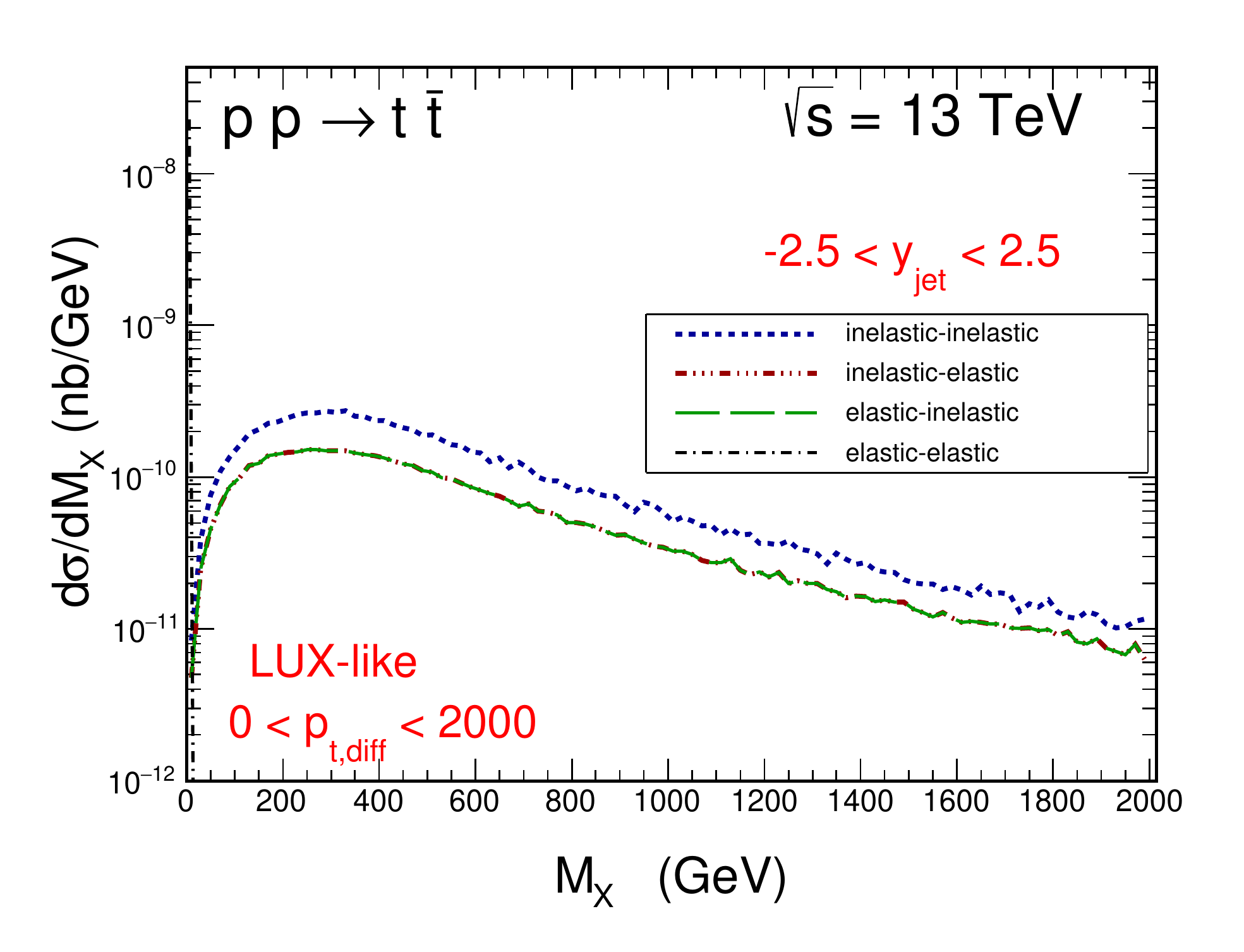}
    \caption{Distribution in the mass of the dissociated system
for different components defined in the figure.
The left panel is without imposing the condition on the struck
quark/antiquark and the right panel includes the condition.
}
\label{fig:dsig_dMX}
\end{figure}

In addition in Fig.\ref{fig:dsig_dMX} we show distributions
in outgoing proton remnant masses $M_X$ and/or $M_Y$. Similar shapes are observed for
single-dissociative and double-dissociative processes.
Population of large $M_X$ or $M_Y$ masses is associated with the
emissions of jets visible in central detectors (i.e. with -2.5 $< y_{\rm jet} <$ 2.5).
We show the modification of the total cross section through such emissions in the right
panel of the figure. The condition on jet rapidity cuts out the high mass part
of distribution in $M_X$ or $M_Y$. A significant correlation
between maximal $M_X$ and maximal $|y_{\rm jet}|$ can be observed.

\begin{figure}
  \includegraphics[width=.60\textwidth]{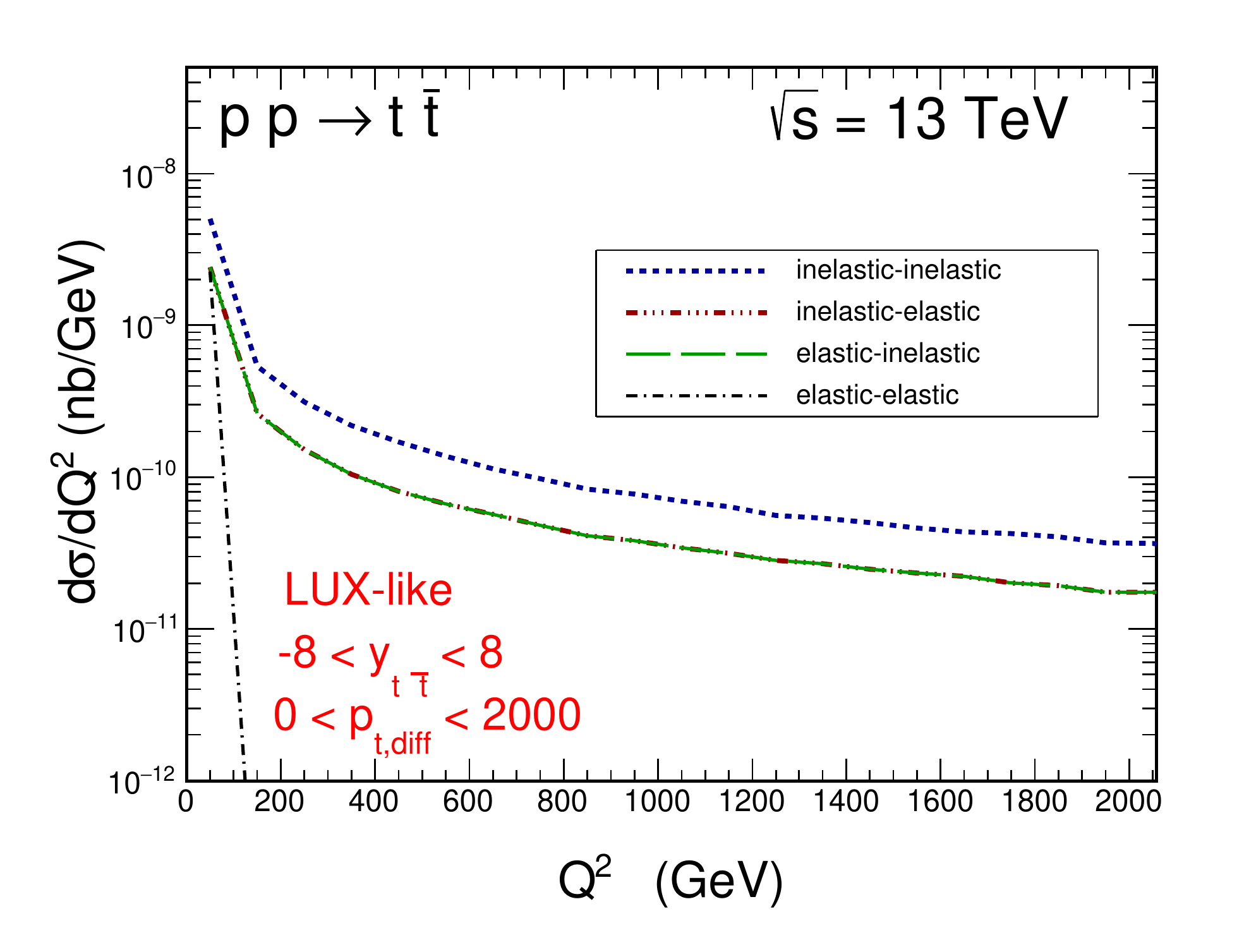}
  \caption{Virtuality distribution
for different components defined in the figure.
}
\label{fig:dsig_dQ2}
\end{figure}

In Fig.\ref{fig:dsig_dQ2} we show distribution in photon virtuality.
One may notice the photon emission through proton dissociation is much broader than that
for elastic production.
A large contribution to the cross section is hence shown to arise from the region
of highly virtual photons, $Q^2 >$ 1000 GeV$^2$.

\begin{figure}
  \includegraphics[width=.60\textwidth]{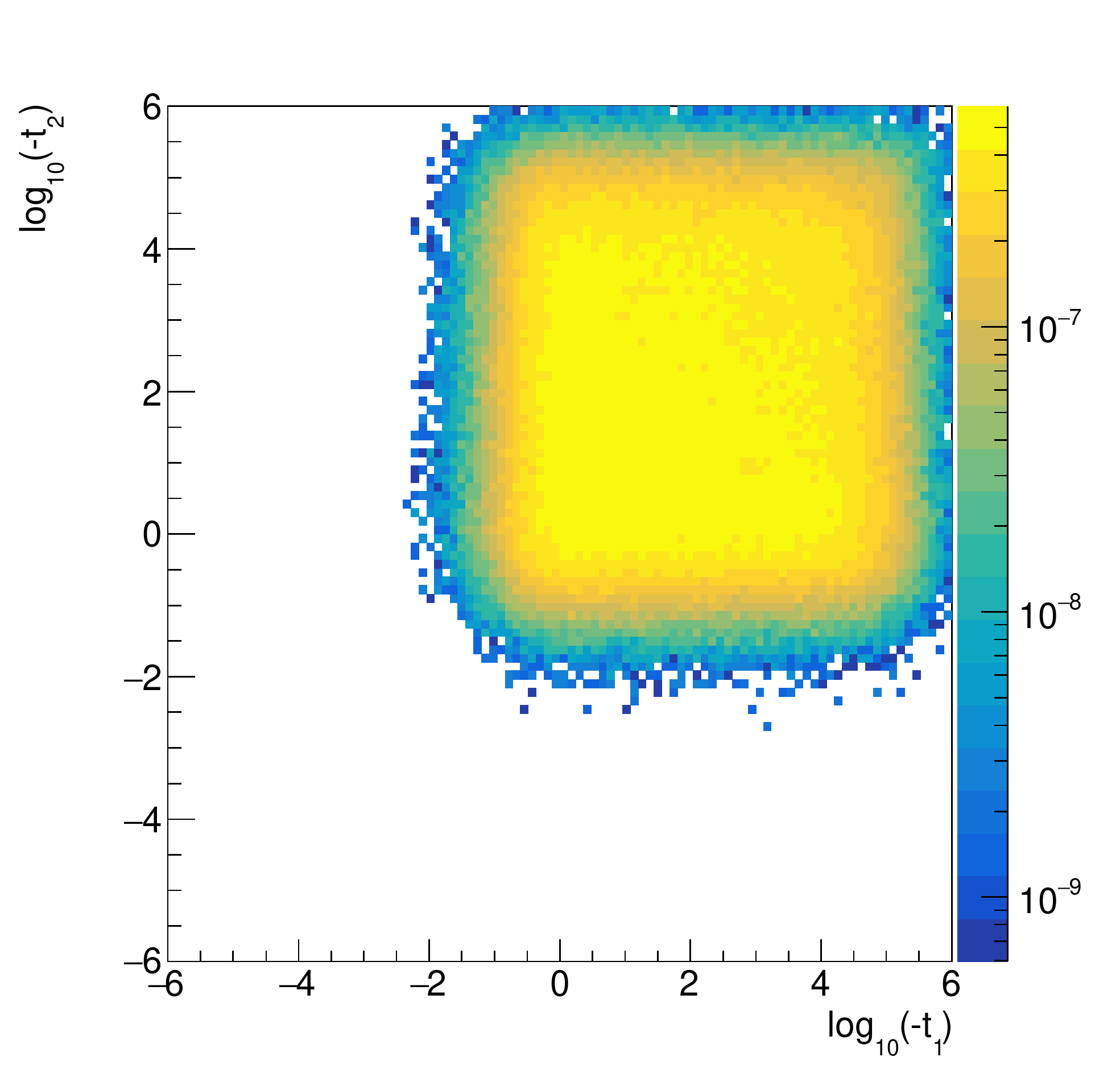}
  \caption{Distribution in $Q_1^2 \times Q_2^2$
for the inelastic-inelastic contribution. Here $Q_i^2 = -t_i$.
}
\label{fig:dsig_dQ12dQ22}
\end{figure}

\begin{figure}
  \includegraphics[width=.60\textwidth]{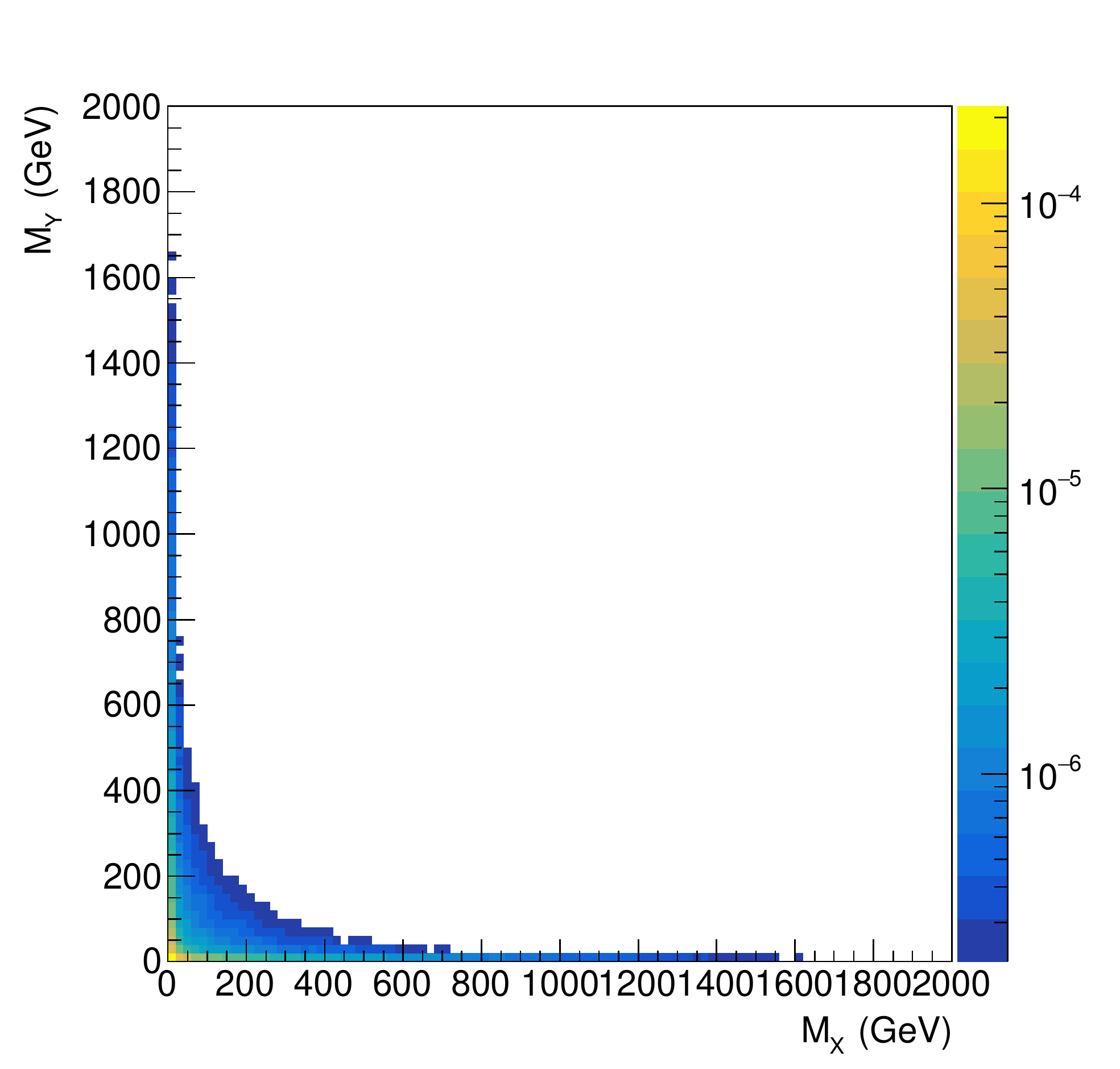}
  \caption{Distribution in $M_X \times M_Y$
for the inelastic-inelastic contribution.
}
\label{fig:dsig_dMXdMY}
\end{figure}

Finally we wish to show some two-dimensional distributions
in $Q_1^2 \times Q_2^2$ (Fig.\ref{fig:dsig_dQ12dQ22}) and
$M_X \times M_Y$
(Fig.\ref{fig:dsig_dMXdMY}) for inelastic-inelastic case.
We see regions when both virtualities are large. The apparent
correlation in $M_X$ and $M_Y$ is only an artifact of the presentation
(logarithmic scale).
In fact there is no such correlation as discussed in our recent study
\cite{Forthomme:2018sxa}.
\begin{figure}
  \includegraphics[width=.60\textwidth]{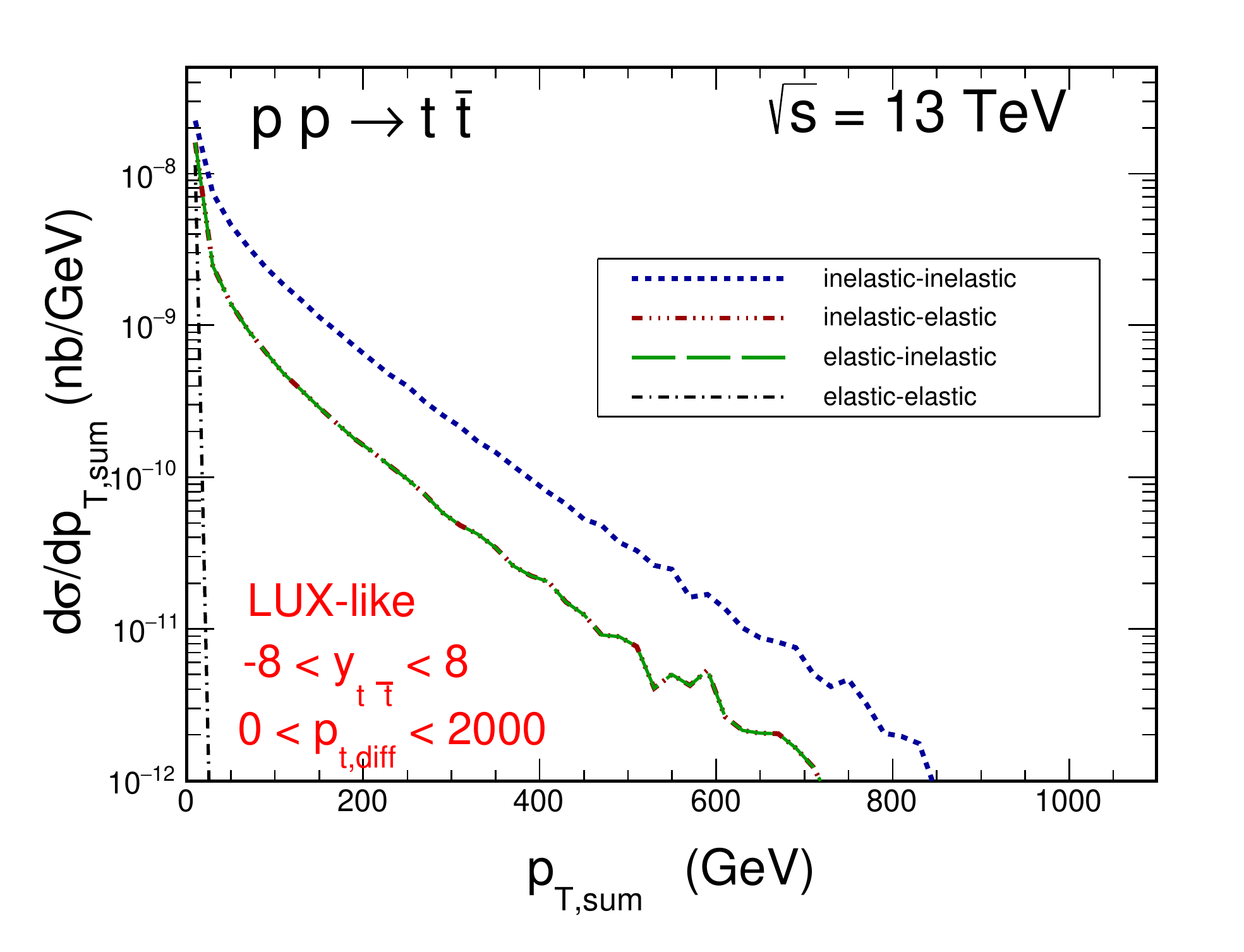}
  \caption{Distribution in transverse momentum of the $t$ and $\bar t$
  pairs for elastic - elastic, inelastic-elastic (elastic-inelastic) and inelastic-inelastic
contributions for LUX-like structure function.
}
\label{fig:dsig_ptsum}
\end{figure}
Finally in Fig.\ref{fig:dsig_ptsum} we show distribution in $p_{\rm T,sum} =
\left|\bp_{\rm T1} + \bp_{\rm T2}\right|$. The distribution for the elastic-elastic component
is much narrower than similar distributions for the other components.
We note that for collinear photon distributions we would obtain a delta-function
in $p_{\rm T,sum}$ for all cases.

\section{The gap survival factor}
Physics of the gap survival due to the jet (parton) emission was discussed
in detail in Ref. \cite{Luszczak:2018ntp} for the $\gamma \gamma \to W^+ W^-$
production.

In this last section we wish to calculate the gap survival
factor at parton level. In such a case the rapidity gap is destroyed by the outgoing parton (jet or mini-jet),
which is struck by the virtual photon.

This gap survival factor may hence be defined as:
\begin{equation}
S_R(\eta_{\rm cut}) = 1 - \frac{1}{\sigma}
\int_{-\eta_{\rm cut}}^{\eta_{\rm cut}} \frac{{\rm d}\sigma}{{\rm d}
    \eta_{\rm jet}} {\rm d} \eta_{\rm jet}, \;
\label{eq:parton_model_gap_survival}
\end{equation}
where ${\rm d}\sigma / {\rm d}\eta_{\rm jet}$ is the rapidity distribution of the cross
section for  $t \bar t$ production as a function of rapidity of the extra
jet (de facto parton) and $\sigma$ is the associated integrated cross
section.
\begin{figure}
  \includegraphics[width=.48\textwidth]{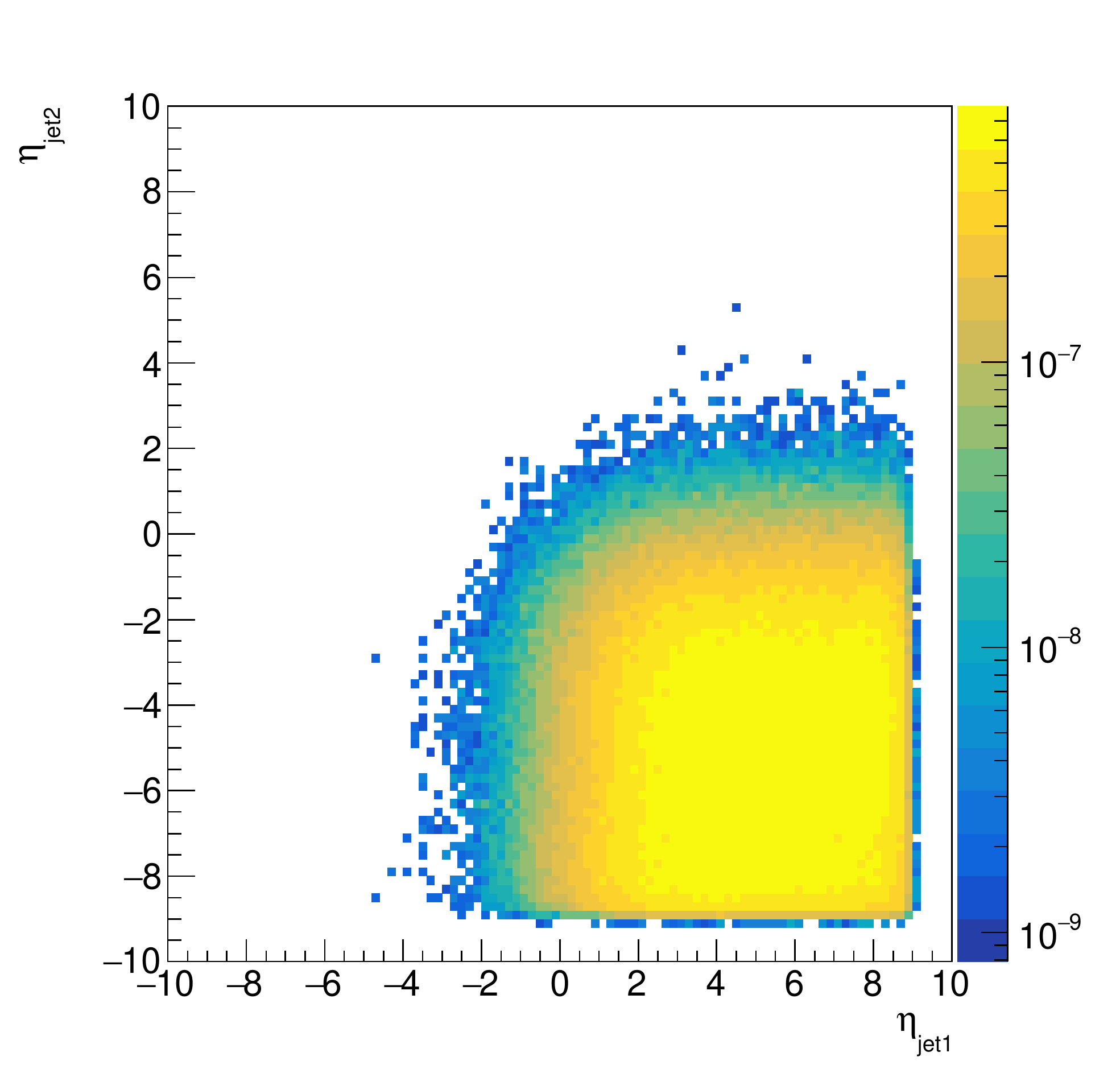}
  \includegraphics[width=.48\textwidth]{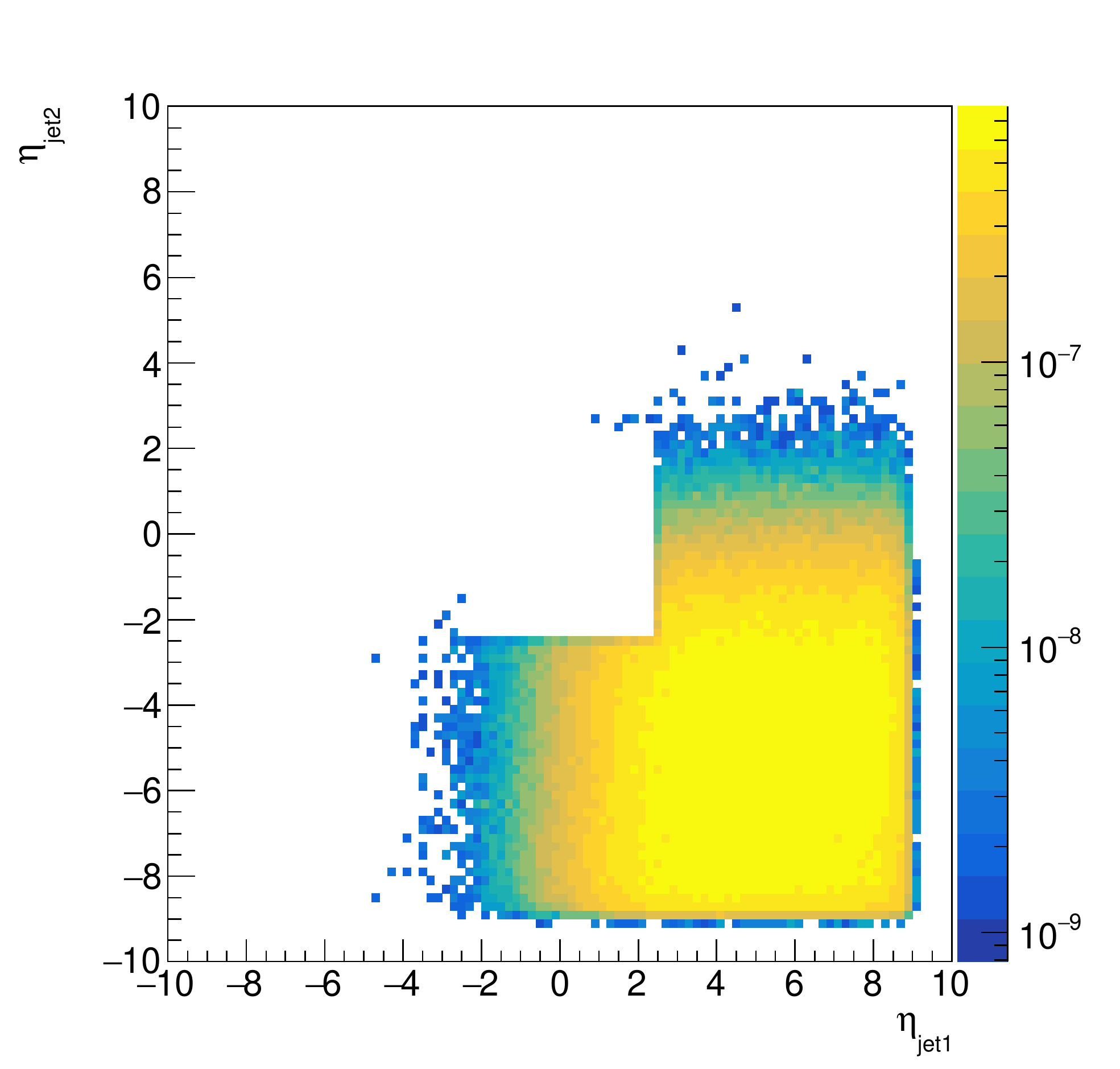}
    \caption{Two-dimensional ($\eta^{\rm ch}_{X},\eta^{\rm ch}_{Y}$) distribution
without cuts (left) and with cuts (right) on $\eta_{jet1}$ and $\eta_{jet2}$.
}
\label{fig:dsig_dy1dy2}
\end{figure}
\begin{figure}
  \includegraphics[width=.50\textwidth]{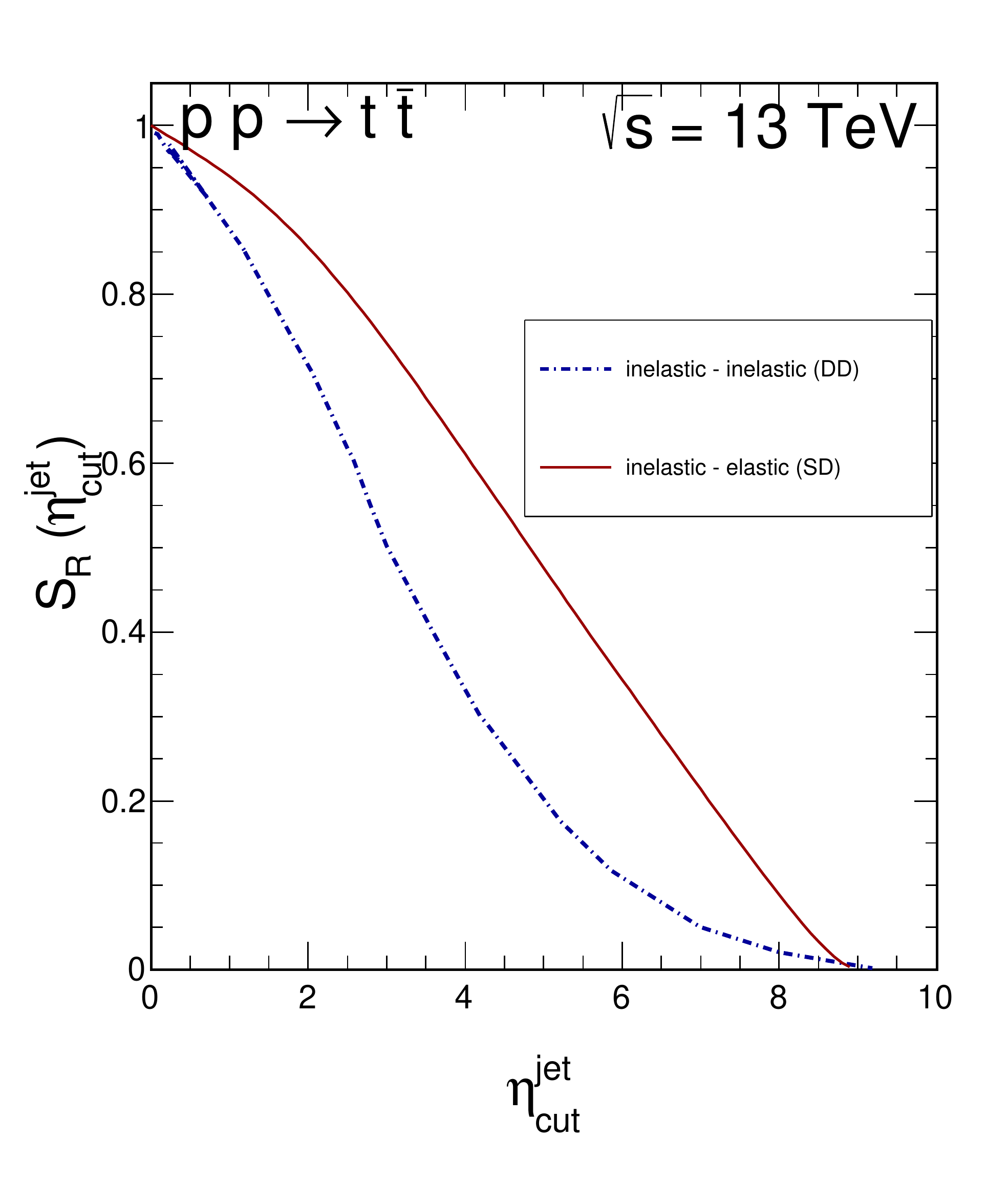}
  \caption{Gap survival factor for single and double dissociation as a function of
    the size of the pseudorapidity veto applied on charged particles emitted from
    proton remnants.
}
\label{fig:s_r_dd}
\end{figure}
Here, in Fig.\ref{fig:s_r_dd} we show our results for
$pp \to \gamma \gamma \to t \bar t$ processes. The gap survival
factor is $S_R^{DD} < S_R^{SD}$. We have checked the factorisation
$S_R^{DD} = (S_R^{SD})^2$. One may note the magnitude of such gap survival factors can reverse the ordering in
\eqref{eq:sigma_hierarchy} for large $\eta_{cut}$.

\section{Conclusions}

In the present paper we have presented cross sections for production of $t \bar t$ pairs via $\gamma^* \gamma^*$
fusion. Such processes can be separated out by imposing rapidity gaps in
the central detector.
Our calculations include transverse momenta of the intermediate
photons.
The flux of photons produced with proton dissociation has been
expressed in terms of proton structure functions. For our study we have used a hybrid
parameterisation of proton structure functions, using similar input
as the recent LUXqed parameterisation \cite{Manohar:2017eqh}.

The cross section summed over the different categories of processes
is about 2.36 fb (full phase space), i.e. rather small compared to
the standard inclusive $t \bar t$ cross section (of the order of nb).
We have shown that the ordering \eqref{eq:sigma_hierarchy} holds for the whole
phase space without extra experimental conditions
on the top quarks decay products reconstruction efficiencies and central system gap survival factor.
As discussed recently in Ref.~\cite{Forthomme:2018sxa} for the $W^+W^-$ final state, the remnant
fragmentation leads to a taming of the cross section when
the rapidity gap requirement is imposed.
Also here such a condition reverses the hierarchy observed for the case
when such condition is taken into account.

One may argue that a diffractive QCD contribution leads to the
same rapidity-gap topology as $\gamma\gamma$-fusion.
We have checked, that using the formalism of \cite{Szczurek:2012mi,Maciula:2010vc},
the cross section for a QCD diffractive
contribution to the $t \bar t$ production is about 2-3 orders of
magnitude smaller than the one corresponding to the photon-photon fusion
and does need to be included in the final estimate.

This is very different than for ``exclusive'' $c \bar c$
\cite{Szczurek:2012mi} or $b \bar b$ \cite{Maciula:2010vc}
production.
We have presented several differential distributions in rapidity and
transverse momentum of the $t$ or $\bar t$ as well as invariant mass
of the $t \bar t$ system or in the mass of the dissociated proton system.
We have shown also some correlation distributions, some of them
two-dimensional ones.

Our results imply that for the production of such heavy objects as $t$ quark
and $\bar t$ antiquark the virtuality of the photons attached to
the dissociative system are very large ($Q^2 <$ 10$^{4}$ GeV$^2$).
A similar effect was discussed in detail already for
the $W^+ W^-$ system \cite{Luszczak:2018ntp}.

We have presented the best estimate of the cross
section(s) and differential distributions for the inclusive case
(no requirement on rapidity gap) as well as including extra condition
on the jet rapidity.
Applying a veto on charged particles or outgoing jet
in a certain rapidity region (as done here) lowers the cross
section significantly.
As the gluon-gluon fusion cross section is so large, rapidity
gap fluctuations in the hadronisation can be a serious background.
The gap must then be chosen to minimise the
unwanted contributions from gluon-gluon and quark-antiquark subprocesses
not loosing too much of the signal ($\gamma \gamma \to t \bar t$ contribution).
Evaluating such an effect will be necessary to demonstrate whether
the Standard Model $\gamma \gamma \to t \bar t$ contribution can
be ``observed'' at the LHC.

\section*{Acknowledgements}
This study was partially supported by the Polish National Science Centre
grants
DEC-2014/15/B/ST2/02528
and by the Center for Innovation and
Transfer of Natural Sciences and Engineering Knowledge in Rzesz{\'o}w.

\bibliographystyle{apsrev}
\bibliography{pp_ttbar}


\end{document}